# Resource Management and Scheduling for Big Data Applications in Cloud Computing Environments

**Muhammed Tawfiqul Islam and Rajkumar Buyya**

Cloud Computing and Distributed Systems (CLOUDS) Laboratory
School of Computing and Information Systems
The University of Melbourne, Australia
**Email: muhammedi@student.unimelb.edu.au, rbuyya@unimelb.edu.au**

This chapter presents software architectures of the big data processing platforms. It also provides in-depth knowledge on resource management techniques involved while deploying big data processing systems in the cloud environment. It starts from the very basics and gradually introduce the core components of resource management which we have divided into multiple layers. It covers the state-of-art practices and researches done in SLA-based resource management with a specific focus on the job scheduling mechanisms.

The key takeaways from this chapter are:

- Basic understanding of cloud computing and big data processing platforms
- Knowledge about the software architectures and use-cases of different big data processing platforms
- Knowing about the vital software and tools for successful deployment and management of big data clusters
- Learning the key steps and significant components of resource management for big data applications through a layered architecture
- Focusing on the existing literature on SLA-based application/job scheduling mechanisms and finding the applicability of these techniques in different application scenarios
- Gaining insights about the research gaps and highlights on the future research challenges

## 1 Introduction

Cloud Computing is an emerging platform which can provide infrastructure, platform, and software for storing and computing of data. Nowadays Cloud Computing is used in many small and large organizations like a utility (Buyya et al. 2009) as it is more affordable to go for the pay per use service of cloud service providers instead buying and maintaining own computing resources. While registering in any Cloud Service, both the cloud service customer and the cloud service provider must agree on some predefined policies which are called the Service Level Agreement (SLA). Violation of SLAs may affect the proper execution and performance of an application of any customer, so it poses a significant threat on a cloud service provider's business reputation. Therefore, it is essential to manage the cloud resources in such a way that it guarantees SLA.

Big Data (Assunção et al. 2015; Kune et al. 2016) is the recent hype in information technology. Scientific applications generate a large amount of data which is used for discoveries and explorations. Besides, social media data analysis, sentiment analysis, and business data analysis are crucial for business organizations to adopt customer needs and gain more profits. Cloud computing can be an appropriate solution to host big data applications, but many challenges need to be addressed to use the existing cloud architectures for big data applications. This chapter discusses the challenges of hosting big data processing platforms in the cloud. Moreover, it also gives a comprehensive overview of cloud resource management for big data applications. Resource management is a broad domain that contains many complex components. However, to make it easier to understand, we divide it as a layered architecture and discuss the critical elements from each layer. Our focus will be on resource allocation and scheduling mechanisms and how the existing research tried to incorporate SLA in these components. We will also point out the limitations of the current approaches and highlight future research directions.

The contents of this chapter are organized as follows. Sections 2 provides background on cloud computing, big data, big data processing platform systems and their architectures and some popular cluster managers. Section 3 gives a layered overview of the overall resource management process for big data applications on the cloud. Section 4 shows a taxonomy of resource allocation for big data applications. Section 5 exhibits a taxonomy of job scheduling mechanisms for big data applications. Section 6 discusses the research gaps and future research directions towards SLA-based resource management. Finally, section 7 concludes the chapter.

## 2  Background

In this section, we briefly discuss the key features of cloud computing. Moreover, we explain the architectures of the popular open-source software systems for processing big data applications. Also, we provide an overview of some popular cluster managers. Finally, we conclude with explaining why the cloud is a viable alternative to deploy a big data processing software and how cluster managers can be used for efficient management of the system.

## 2.1 Cloud Computing

Cloud computing delivers a shared pool of on-demand computing resources on a pay-per-use basis. The main features of cloud computing are:

- **Resource elasticity:** cloud resources can be easily scaled up or down to meet application or user demands.

- **Metered service:** users are billed based on what resources they used and how long they have used them.

- **Easy access:** the resources of cloud can be easily accessed and can be provisioned as a self-service manner.

There are three different types of cloud. These are:

- **Public Cloud:** There are many public cloud service providers who offers computing resources as a pay-per-use basis. Organizations can hire resources from these service providers to deploy their own applications. It greatly reduces the cost of buying computing hardware and removes the burdens of managing local resources.

- **Private Cloud:** Many organizations setup an on-premise computing resource facility which is known as the private cloud. The main reason for setting up a private cloud is to reduce the data transfer overhead to the public cloud. In addition, it also ensures that private and sensitive data are kept on the organization's premises to reduce security threats.

- **Hybrid Cloud:** It is mix of both public and private cloud. Sometimes organizations need to scale up their resources in public cloud to be processed in the public cloud.

Cloud provides computing as a service, and we can divide cloud services in three ways:

- **Software as a Service (SaaS):** SaaS can be used from any devices through the Internet and typically these services are accessed via a web browser. The required software needed by the user for any specialized task are already developed and provided thorough different interfaces. Users just define the task, input data and collect the results. Example: Google Apps[1]

---

[1] https://gsuite.google.com.au/intl/en_au/

- **Platform as a Service (PaaS):** A platform is provided for developing distributed, scalable cloud-based programs. It greatly reduces the hassle for managing the underlying resources. Example: IBM Cloud[2]

- **Infrastructure as a Service (IaaS):** Computing and storage resources are provided to setup a user's own infrastructure to build platform and services. Reduces the hassle of buying and managing own physical hardware, provides a scalable on demand pool of resources. Example: Windows Azure[3], Amazon EC2[4].

## 2.2 Big Data

In today's world, huge amount of data is being generated through social media, scientific explorations and many other emerging applications like Internet of Things (IoT). The term "Big data" is not about the size of data; rather it covers many other aspects. For simplicity, we can define it in terms of the 3V as shown in Figure 1.

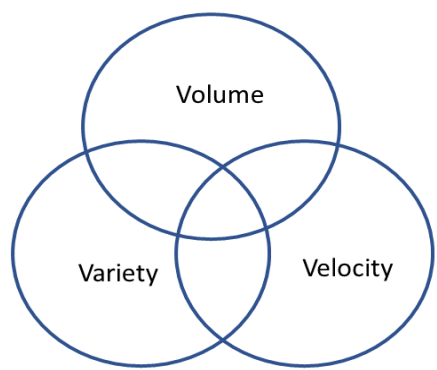

Figure 1: Big Data 3V

The volume of data can be small or large, from a few Megabytes to thousands of Terabytes. Each day we are generating so much data that recently (in 2015) we have moved into a Zettabyte era. Velocity represents the speed of incoming data. For example, some applications need real-time or near real-time processing and comes with great speed. These types of applications can be categorized as streaming applications. In contrast, applications that need offline processing of huge volume of static data are called batch applications. Finally,

---

data can have many varieties such as structured, unstructured etc. Storing and processing of data is often not possible by the traditional Database Management Systems (DBMS) and NoSQL has greatly replaced SQL in many domains. There are many other aspects of big data (many other Vs) depending on the specific domain.

## 2.3 Big Data Processing Platforms

Processing big data is a difficult task, and it is not possible in a centralized system. Therefore, distributed computing solutions are used for parallel processing of big data. Many big data processing platforms have emerged over the last decade. Figure 2 shows a taxonomy on big data processing platforms. As it shows, previously only batch-based platforms like Hadoop was mostly used. However, due to the discovery of many scientific, business and social streaming applications, real-time processing became more influential and dedicated stream processing platforms like Strom, S4 were invented. However, applications became more complex, and often organizations need to have both batch and stream-based processing. Hence, some hybrid processing platforms like Apache Spark, Apache Flink are being used in the industry.

In this chapter, we only focus on batch and hybrid-based processing platforms and briefly discuss about some of the most popular ones.

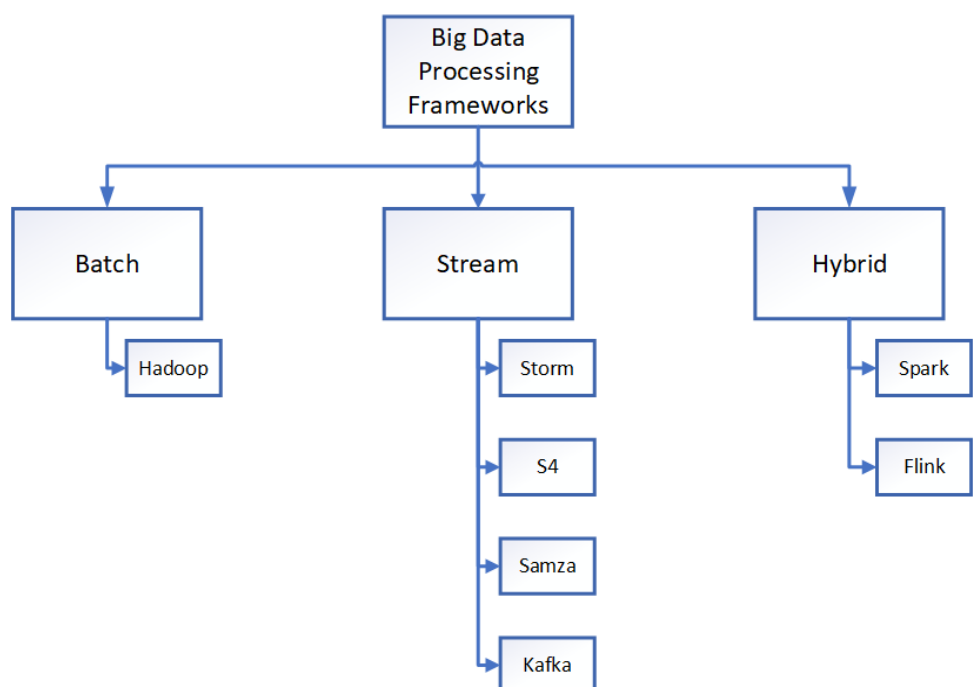

Figure 2: A Taxonomy of Big Data Processing Frameworks

### 2.3.1  Apache Hadoop

Apache Hadoop, introduced by Yahoo in 2005, is the open source implementation of the MapReduce programming paradigm. The main feature of Hadoop is to use primarily distributed commodity hardware to parallel processing of batch-based jobs. The core of Hadoop is its fault-tolerant file system Hadoop Distributed File Systems (HDFS) (Shvachko et al. 2013) that can be explicitly defined to span in many computers. In HDFS, the block of data is much larger than a traditional file system (4KB versus 128MB). Therefore, it reduces the memory needed to store the metadata on data block locations. Besides, it reduces the seek operation in big files. Furthermore, it greatly enhances the use of the network as only a fewer number of network connections are needed for shuffle operations. In the architecture of HDFS, there are mainly two types of nodes: Name node and Data node. Name node contains the metadata of the HDFS blocks, and the data node is the location where the actual data is stored. By default, three copies of the same block are stored over the data nodes to make the system fault tolerant. The resource manager of Hadoop is called Yarn (Vavilapalli et al. 2013). It is composed of a central Resource Manager who resides in the master node and many Node Managers that live on the slave nodes. When an application is submitted to the cluster, the Application Master negotiates resources with the Resource Manager and starts container (where actual processing is done) on the slave nodes.

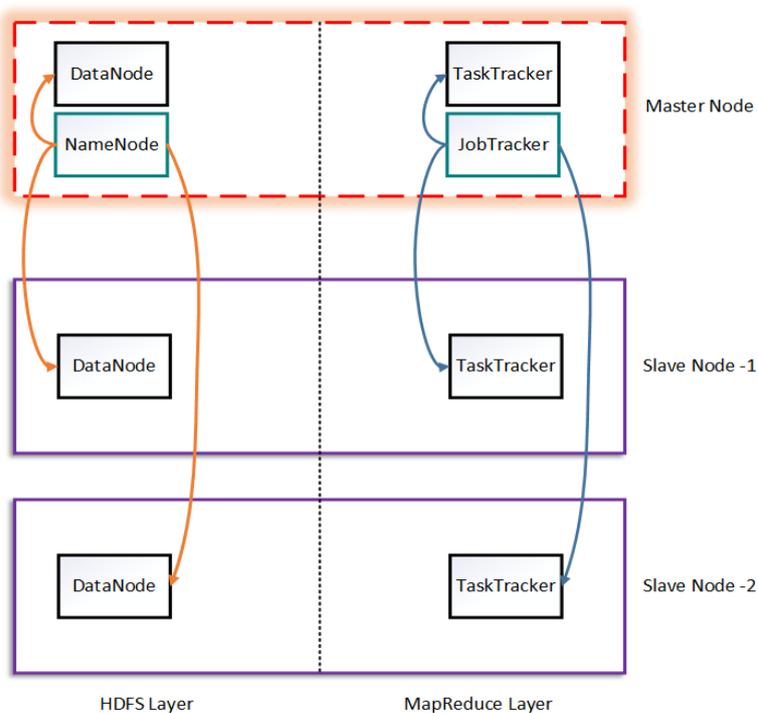

Figure 3: Apache Hadoop Architecture

The main drawback of Hadoop was that it stored intermediate results in the disk, so for shuffle-intensive operations like iterative machine learning, a tremendous amount of data is stored in the disk and transferred over the network which poses a significant overhead on the whole system.

### 2.3.2 Apache Spark

Apache Spark (Zaharia et al. 2016) is one of the most prominent big data processing platforms. It is an open source, general-purpose, large-scale data processing framework. It mainly focuses on high-speed cluster computing and provides extensible and interactive analysis through high-level APIs. Spark supports batch or stream data analytics, machine learning and graph processing. It can also access diverse data sources like HDFS, HBase (George 2011), Cassandra (Lakshman and Malik 2010), etc. and use Resilient Distributed Dataset (RDD) (Zaharia et al. 2012) for data abstraction.

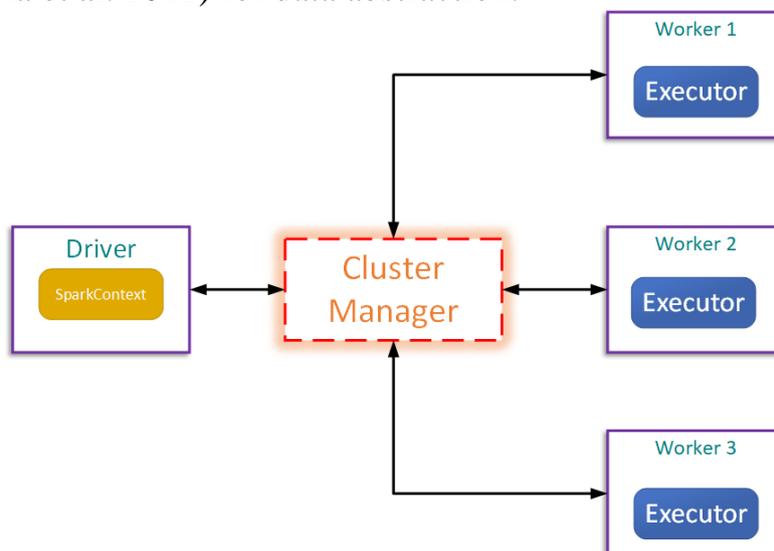

Figure 4: Apache Spark Architecture

As compared to the Hadoop system tasks, Apache Spark allows most of the computations to be performed in memory and provides better performance for some applications such as iterative algorithms. When the results do not fit on the memory, the intermediate results are written to the disk. Spark can run locally in a single desktop, in a local cluster, and on the cloud. It runs on top of Hadoop Yarn, Apache Mesos (Hindman et al. 2011) and the default standalone cluster manager. Jobs/applications are divided into multiple sets of tasks called stages which are inter-dependent. All these stages make a directed acyclic graph (DAG), where each stage is executed one after another.

### 2.3.3 Apache Flink

Apache Flink (Katsifodimos and Schelter 2016) is an open-source stream processing platform. It executes data-flow programs in data-parallel pipelines. Flink is fault-tolerant and treats batch data as a form of a stream, therefore, it is a hybrid framework. Programs can be written in Java, Scala, Python, and SQL. Flink does not provide any data storage mechanism. Instead, it uses other data sources like HDFS, Cassandra, etc. During the execution stage, Flink programs are mapped to streaming dataflows. Every dataflow starts with one or more origins (input, queue or file system) and ends with one or more sinks (output, message queue, database or file system). An arbitrary number of transformations can be done on the stream. These dataflow streams are arranged as a directed acyclic dataflow graph, allowing the flexibility for the applications to branch and merge dataflows. Flink is relatively new and unstable as compared to the matured frameworks like Hadoop and Spark. It is yet to be seen whether Flink can be scalable like Spark in a production-grade cluster.

## 2.4  Cluster Managers

### 2.4.1  Apache Hadoop Yarn

Apache Hadoop Yarn (Vavilapalli et al. 2013) is the resource manager for Apache Hadoop. The core idea of Yarn is to split up the mechanisms for resource management such as job scheduling, monitoring, etc. into separate daemons. There is a global Resource-Manager in the master node and Node Managers in each of the worker/slave nodes. Resource Managers and Node Managers form the whole data-computation framework. Resource Manager is the ultimate co-ordinate that can dictate resource provisioning and scheduling in the entire system. Node Managers are responsible for running containers and monitor resource usages and reporting the resource usage statistics to the Resource Manager. Furthermore, per-application Application-Manager negotiates with the Resource Manager to reserve resources and collaborates with the Node Manager to run containers and monitor the tasks.

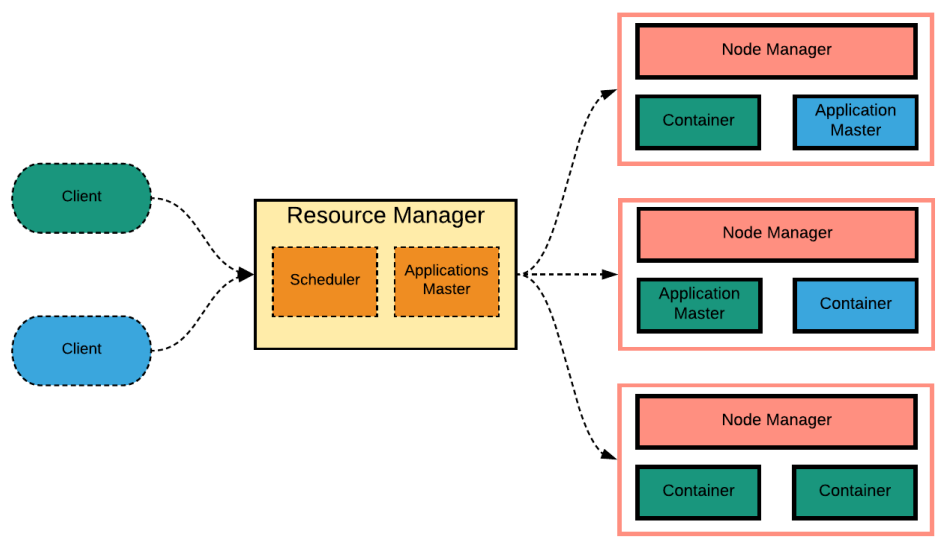

Figure 5: Apache Hadoop Yarn Architecture

The Resource Manager has two main components: Scheduler and Applications-Manager. Scheduler tracks and maintains a queue of jobs set the order of the jobs and allocate resources to each of the jobs before execution. The scheduler functions are based on the implemented policies and SLA requirements of the applications. The scheduler has a pluggable policy which makes it extendable to different scheduling policies. For example, CapacityScheduler and FairScheduler are the example plugins implemented and available with Yarn. Applications-Manager accepts job submission requests and provides the service to restart failed jobs.

### 2.4.2  Apache Mesos

Apache Mesos (Hindman et al. 2011) is said to be the data-center level cluster manager. Mesos was built primarily to support multiple different big data processing frameworks to be running in the same cluster. Mesos isolates the resources (e.g., CPU, Memory and disk) shared by different framework tasks/executors and run them in the same physical/ virtual machine. Schedulers from different frameworks negotiate with Mesos to reserve resources for running tasks. Moreover, each application (of any big data processing system like Spark, Hadoop, Storm) is called a framework and can have a custom implemented scheduler that can negotiate with Mesos to set the required resources for that application.

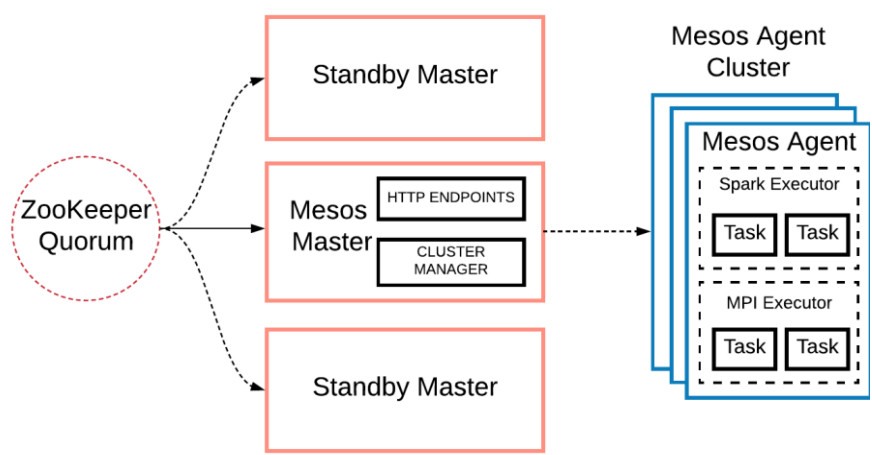

Figure 6: Apache Mesos Architecture

Mesos send resource offers to each framework by using the Dominant Resource Fairness (DRF) (Ghodsi et al. 2011) resource allocator which tries to distribute the resources among multiple frameworks equitably. However, Mesos has an advanced scheduler and operator HTTP APIs and supports Dynamic

Resource Reservation for any application. Therefore, by using the scheduler/operator APIs, it is possible to build custom pluggable scheduler with specific SLA requirements. Frameworks can also be assigned with particular roles and set resource quotas to make the resource management flexible.

### 2.4.3 Google Kubernetes

Kubernetes[5] is an open source container management platform which is designed to run at production scale. It was built upon the foundations laid by Google. The architecture of Kubernetes supports loosely-couped mechanism for service discovery. There are a master and one or more computing nodes in a Kubernetes cluster. The master exposes APIs, schedules workloads and controls the cluster. Each node runs a container runtime like Docker or rkt an agent that communicates with the master. A node also has additional components responsible for logging, monitoring, service discovery and optional add-ons. A pod is a collection of containers that serve as a core unit of management. It acts as logical isolation for containers sharing same context and resources. Replica sets provide the required scale and availability of services by maintaining a pre-defined set of pods. The deployment of an application can be scaled by using replica sets which ensures an application has its desired number of pods running to meet the requirements. The master node has etcd, which is an open-source distributed key-value database and acts as the single point of truth for all components in a Kubernetes cluster. When an application gets enough pods to run, the nodes pull images from the image registry and works with the local container runtime to launch the container in each pod. Kubernetes is flexible and provides a rich set of APIs for building custom container management modules which are particularly useful in deploying efficient, large-scale IoT/Fog based applications.

## 3 Resource Management for Big Data Applications

In this section, we will provide a brief overview of the significant components of resource management for Big Data applications. Many steps or components can be included. However, the overall process of managing resources for big data applications is a complex task, and many parts are inter-dependent thus it is hard to distinguish them. Therefore, as shown in Figure 3, we have simplified the categorization in three different layers and only discuss the key elements from each of these categories.

---

[5]     https://kubernetes.io/

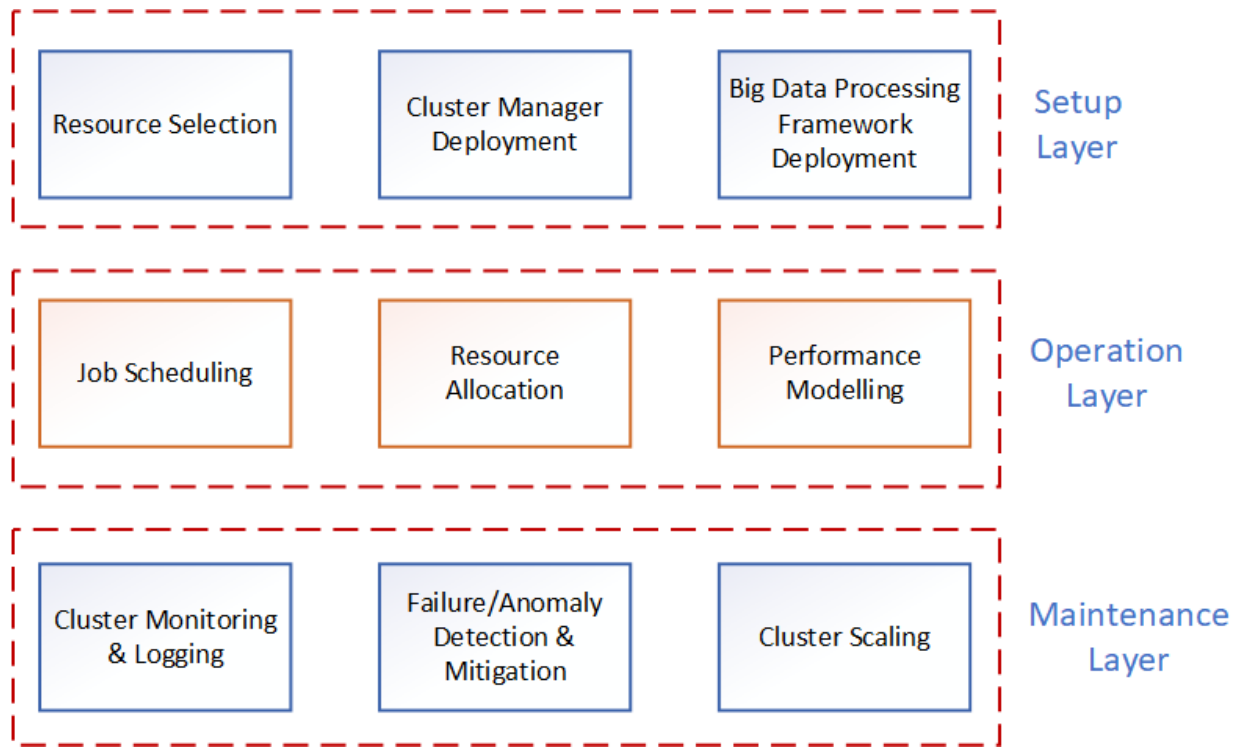

Figure 8: Key components of Resource Management in a Big Data Cluster

❖ Resource Management Layers:

## 3.1 Setup Layer

The first layer of resource management is the Cluster Setup. In this layer, hardware or virtualized resources are selected depending on the applications. Additionally, a cluster manager is deployed to manage the resources and jobs from different big data processing frameworks. Lastly, one or more big data processing frameworks are used.

### 3.1.1 Resource Selection

Both physical or virtualized resources can be used to build a cluster. Generally, depending on the applications and analytics demands of any business organization, the hardware resources are chosen. The setup can be done on-premise (local cluster or private cloud), deployed on cloud resources (public cloud) or a hybrid deployment (some local resources with a pay-as-you-go subscription from a cloud provider) can also be made. The actual underlying hardware resources might vary with applications. However, CPU, RAM, Storage, and Network are the must no matter where the cluster is deployed. Nowadays, GPU resources are gaining popularity due to the widespread use in sophisticated machine learning (deep learning) algorithms running in platforms like TensorFlow.

### 3.1.2  Cluster Manager Deployment

The next step is to choose a cluster manager to manage both the jobs and the resources. A cluster manager also balances the workloads and resource shares in a multi-tenant environment. For containerized applications, Kubernetes or Docker Swarm can be deployed to provide container management platform. Kubernetes excels as a complete management system featuring scheduling, dynamic on-the-fly updates, auto-scaling, and health monitoring. However, Docker Swarm features a system-wide view of the whole cluster from a single Docker engine. Apache Hadoop Yarn is the cluster manager of choice if all the applications of the cluster are only MapReduce or Hadoop-based. In contrast, Apache Mesos is a better choice than Yarn as it supports efficient resource isolations for multiple different big data processing frameworks and provides strong scheduling capabilities.

### 3.1.3  Big Data Processing Framework Deployment

Many big data processing frameworks are available which can run distributed applications across one or more clusters. The applications can be real-time, stream or batch and for each type of applications, there are some frameworks which are capable of handling the requirements efficiently. It is not possible to say which is the best possible framework to deploy in general. Instead, each one has its own merits and suits a group of applications. For example, in the last decade, Hadoop was the most prominent platform to process MapReduce based static batch jobs. However, due to the increasing popularity of real-time systems and streaming applications; Apache Spark, Apache Flink, and Apache Storm have become the standard choice to tackle them. Apache Storm is particularly useful for stream-based applications. Apache Spark is vastly replacing both Hadoop and Strom, and it is a hybrid platform that supports both batch and stream processing. Apache Flink is new a hybrid platform that needs to be more stable to compete with the likes of Spark or Storm.

## 3.2   Operation Layer

The second layer of resource management is the operation layer. Here, performance models are built to determine the set of resources to be allocated that is enough to meet user SLA and schedule multiple jobs in a multi-tenant setup. Moreover, the overall cluster utilization is maximized, and each job's performance is enhanced without interfering with any other job's SLA.

### 3.2.1   Performance Modelling

The performance of a job might vary depending on various aspects like allocated resources, workload size, task placement, etc. Hence, before a complete deployment of a job, performance models can be established which will be used in resource allocation and job scheduling phase to choose an optimized set of resources to run the job without sacrificing any performance constraints. Generally, performance modeling can be done in two ways. First, running the job with different resource configurations and workloads to build job profiles. Second, collecting historical data of jobs running in the cluster. Both job profiles or historical data can be used to perform statistical analyses, training machine learning algorithms or build mathematical models. These models are then used to select optimal resource configurations and efficient scheduling strategies. There are many existing researches that tried to model the performance of different types of jobs running in both Spark or Hadoop based platforms. (Zhuoyao Zhang, Cherkasova, and Loo 2013) modeled the performance of MapReduce workloads in a heterogeneous cluster (where resources are different types, or the performance varies). This model is then used to predict the job completion times. (Wang and Khan 2015) proposed a simulation-based approach where they have used different Apache Spark configuration parameters and modeled different stages of a job to predict its completion time.

### 3.2.2   Resource Allocation

Resource allocation means reserving a set of resources for a job which will be used by that job to run its tasks up to a specific period. Generally, resource allocation is of two types.

•     Static: Manual resource allocation for each job by the user if the user has enough knowledge on the application behavior on the cluster environment.

• Dynamic: The job is started with a few sets of resources. Based on the utilization and to meet the SLA constraints, more resources might be allocated or deallocated over time.

Choosing the right amount of resources to meet user SLA is crucial as improper resource allocation might lead to either under-utilization or over-utilization problem. Therefore, as mentioned in the previous step, performance models are used to determine the optimal set of resources for each job. Resource allocation can be done from both big data processing framework or cluster manager side. (Islam, Karunasekera, and Buyya 2017) modeled Spark jobs based on different parameters such as input size, iteration, resource requirements to predict the job runtime. Then an optimized resource configuration parameter is suggested based on the models which is enough for that job to meet its deadline. (Sidhanta, Golab, and Mukhopadhyay 2016) also suggested a deadline-aware model to perform resource allocation which is also cost-effective. The model is called OptEx and it estimates job runtime before resource allocation by using the job profiles. (Verma, Cherkasova, and Campbell 2011) proposed a resource provisioning framework for MapReduce jobs which also uses job profiles from jobs to estimate the required resources for jobs.

### 3.2.3 Job Scheduling

It is the most critical component of resource management. Job scheduling means settings the order of the jobs in which they will run on the cluster. Additionally, the resources can also be ordered before running any jobs. Both job and resource ordering depend on the scheduling policy. The most straightforward scheduling policy that is used in all the cluster managers and big data frameworks is the FIFO (First in First Out). Here, jobs are ordered according to their arrival time; that means the job that comes first is executed first in the cluster. If there are not enough resources in the cluster to run all the jobs, then the remaining jobs are placed in a queue which is sorted based in increasing order of their arrival time. In most cases, the FIFO scheduler underperforms with complex SLA requirements in a multi-tenant cluster setup. Therefore, a vast amount of research exists in this area that proposes efficient schedulers with optimal scheduling policies. However, most of the scheduling algorithms are either application or the SLA-demand specific. Also, the parameters that are considered vary greatly depending on the application or cluster setting. The more sophisticated schedulers tackle both resource allocation and scheduling together to make it more efficient. First, these schedulers use some pre-existed performance models for the jobs at hand or build it dynamically then decide the resource configuration for a job before scheduling it. Moreover, the resource usages of the currently executing jobs are tracked, and further resources will be reallocated or deallocated to make it optimally achieve the SLA requirements. Job scheduling is a massively broad and explored topic in both big data and cloud computing. We will provide a detailed discussion and compare the existing works in section 4.

## 3.3  Maintenance Layer

It is the final layer of resource management for big data applications. The components of this layer are responsible for maintaining an already deployed big data cluster.

### 3.3.1  Cluster Monitoring and Logging

Cluster Monitoring is crucial as it plays a vital role in the resource management lifecycle. The cluster monitoring data can be logged and saved in persistent storage. This data can be used to validate the performance of the resource allocation and scheduling policies. Besides, if a feedback-based system is used (can be both feedback-based resource reprovisioning/ scheduling and machine learning models that are updated and improved by using the current system status), it needs to use the cluster monitoring data to improve the system performance. Popular big data processing frameworks like Hadoop, Spark, and Storm provide cluster-wide monitoring data and web-UI to visualize the health of the cluster. Besides, cluster monitoring data can also be found from cluster/container management systems like Kubernetes, Mesos and Yarn. Sometimes while building sophisticated application/user-specific resource management modules, data from the underlying platform might not be enough. In those cases, the administrator or developer might need fine-grained resource usage and health data which is possible by using tools such as Collectd[6] or Prometheus[7]. A cluster monitoring system like Prometheus not only provides cluster monitoring data, but it can also offer a time-series database to store the monitoring data. The database is particularly useful for applying advanced machine learning algorithms or performing time-series analysis on the monitoring data.

---

[6]    https://collectd.org/
[7]    https://prometheus.io/

### 3.3.2  Failure/Anomaly Detection & Mitigation

When a cluster is deployed, and in operation, jobs might fail due to an anomaly in the system, hardware/software failure, resource over-utilization, resource-scarcity, etc. By analyzing cluster-wide monitored log data, it is possible to detect the root cause of failures in the system. It is important to solve the issue to keep the cluster healthy so that the jobs can meet their SLAs. The most trivial way to solve the failure is to restart the failed jobs. In the case of resource scarcity, jobs might fail due to a shortage of resource or interference of co-located jobs. This problem can be solved by throwing more resources in the cluster so that the jobs can run properly. In case of hardware or software failures, the affected hardware that might be prone to failure can be avoided in scheduling to avoid any further failures. Chronos (Yildiz et al. 2015) is a Hadoop-based failure-aware scheduler that uses pre-emption on failed jobs. Then it recovers from failure by reallocating the failed jobs with pre-empted resources to meet the SLA objectives. Fuxi (Zhuo Zhang et al. 2014) is fault-tolerant resource management and scheduling system that can predict and prevent failures in large clusters to satisfy user performance needs.

### 3.3.3  Cluster Scaling

A big data processing cluster might need to be scaled up and down based on the current usage. In a high-load hour, the currently running VMs might not be enough to run all the jobs while satisfying all the users' SLAs. Therefore, in this situation, the cluster needs to be scaled up to satisfy the peak surge of resource demands. In contrast, in a light-load hour, a cluster might go under-utilized. In this scenario, the existing cluster jobs can be consolidated in fewer VMs so that the underutilized VMs can be freed and turned off. Dynamically scaling up or down the cluster is possible by using elastic cloud services offered by Amazon AWS or Azure. (Gandhi et al. 2016) is a model-driven autoscaler for Hadoop clusters. It uses novel gray-box performance models to predict job runtimes and resource requirements to dynamically scale the cluster so that SLA is satisfied.

# 4 A Taxonomy on Scheduling of Big Data Applications

Many types of research have been done in the task and resource scheduling in the cloud computing environment. Researchers are trying to adapt existing scheduling approaches to facilitate the needs of big data applications. However, many challenges are posed due to the different characteristics of big data applications. In this section, different scheduling policies will be discussed. We have divided job scheduling approaches for big data applications based on four aspects. Figure 9 exhibits a taxonomy of big data job scheduling in the cloud.

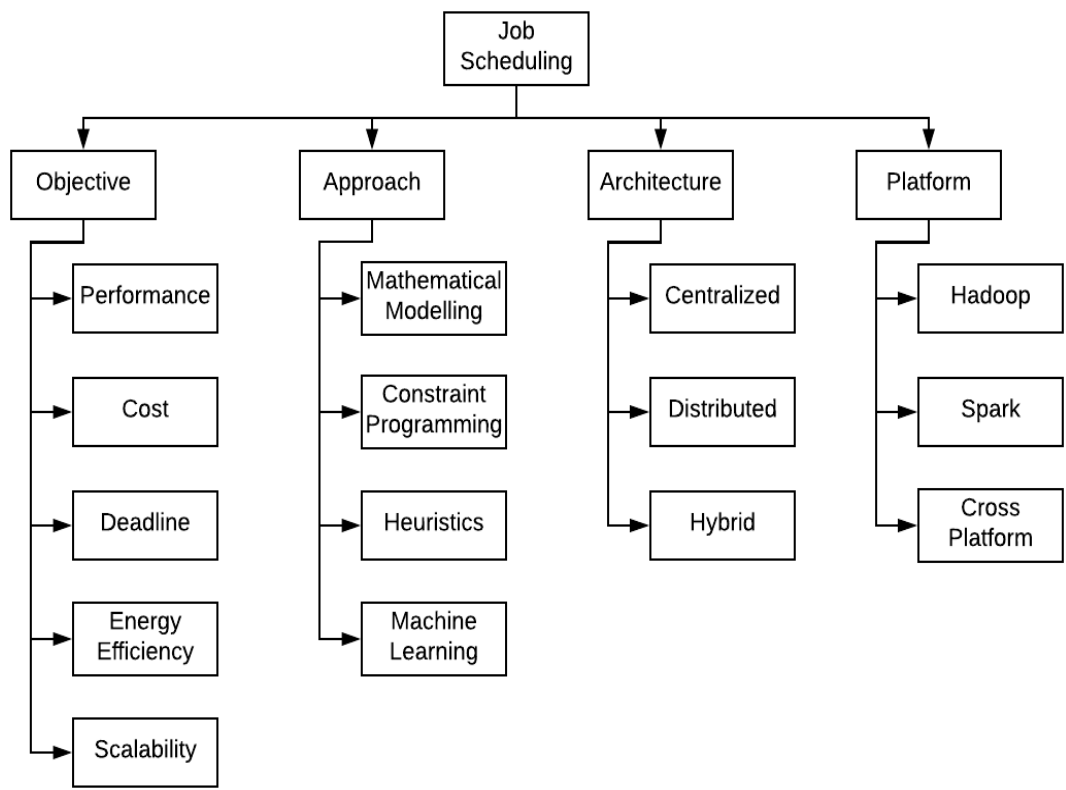

Figure 9: A Taxonomy of Scheduling of Big Data Applications on Cloud

Based on the taxonomy, Table 1 shows a summary of comparison between the existing studies on Job scheduling for big data. In the following subsections, a detailed comparison will be provided between all these works regarding the critical aspects of scheduling. When referring to a paper, we will follow the serial number of the corresponding paper from Table 1.

## 4.1 Objective

The target of a scheduling algorithm to achieve is called the objective. A scheduler can be single-objective or multi-objective, and it depends on the application scenario. Most of the scheduling algorithms focus on improving application performance. Besides, monetary cost reduction, handling soft or tight deadlines of jobs, energy-efficient placement of jobs and scalability of the overall system are also important objectives. Generally, the more objective is added to a scheduler, the more complex the decision-making progress becomes. Sometimes, the overhead of the scheduling solution could be a bigger issue rather than achieving the objects. Therefore, in real systems, different trade-offs are made on the objectives to design fast schedulers with fewer overheads.

Table 1: Comparison between the existing scheduling algorithms

| SL No. | Literature | Objective | Approach | Architecture | Platform |
|---|---|---|---|---|---|
| 1 | (Mashayekhy et al. 2015) | Deadline, Cost, Energy-efficiency | ILP, Heuristic | Centralized | Cross-Platform |
| 2 | (Ousterhout et al. 2013) | Performance, Scalability | Sampling, Late bind | Distributed | Spark |
| 3 | (Ren et al. 2015) | Performance, Scalability | Sampling, SRPT | Hybrid | Cross-Platform |
| 4 | (Sandhu and Sood 2015) | Cost, Scalability | AKNN, Naive Bayes | Hybrid | Cross-Platform |
| 5 | (Zhao et al. 2015) | Deadline, Cost | Greedy Heuristics, ILP | Centralized | Cross-Platform |
| 6 | (Kaur and Chana 2014) | Cost, Scalability | Prediction-based | Centralized | |
| 7 | (Alrokayan, Vahid Dastjerdi, and Buyya 2015) | Deadline, Cost | Prune Tree, Greedy Heuristics | Centralized | Cross-Platform |
| 8 | (Lim, Majumdar, and Ashwood-Smith 2014) | Performance, Deadline | Constraint Programming | Centralized | Hadoop |
| 9 | (Maroulis, Zacheilas, and Kalogeraki 2017a) | Deadline, Energy-efficiency | EDF, Periodic-DVFS | Centralized | Spark |
| 10 | (Lu et al. 2016) | Performance | Genetic algorithm | Centralized | Hadoop |
| 11 | (Rasooli and Down 2012) | Performance, Scalability | Multiplexing | Hybrid | Hadoop |
| 12 | (Fonseca Reyna et al. 2015) | Performance | Reinforcement learning | Centralized | Hadoop |
| 13 | (Nayak et al. 2015) | Performance, Deadline | Greedy, Negotiation | Centralized | Hadoop |
| 14 | (Zacheilas and Kalogeraki 2016) | Performance, Deadline, Cost | Pareto-Frontier | Centralized | Hadoop |
| 15 | (Yildiz et al. 2015) | Performance | Task pre-emption | Centralized | Hadoop |
| 16 | (Zeng et al. 2017) | Performance, Cost, Deadline | Greedy Heuristics | Centralized | Hadoop |
| 17 | (Sidhanta, Golab, and Mukhopadhyay 2016) | Performance, Cost, Deadline | Mathematical model, Prediction | Centralized | Spark |
| 18 | (Cheng et al. 2017) | Performance | Reservation aware, Dependency-aware | Centralized | Cross-Platform |

| 19 | (Chen, Lin, and Kuo 2014) | Performance, Deadline | Graph Modelling | Centralized | Hadoop |
|----|----|----|----|----|----|
| 20 | (Kim et al. 2016) | Cost, Energy-efficiency | Reinforcement Learning | Centralized | Cross-Platform |
| 21 | (Imes, Hofmeyr, and Hoffmann 2018) | Performance, Energy-efficiency | Machine Learning Classifiers | Centralized | Hadoop |
| 22 | (F. Zhang et al. 2014) | Performance, Cost | Evolutionary algorithm | Centralized | Cross-Platform |
| 23 | (Maroulis, Zacheilas, and Kalogeraki 2017b) | Performance, Energy-efficiency | Time-series prediction, DVFS | Centralized | Spark |
| 24 | (Zong, Ge, and Gu 2017) | Performance, Energy-efficiency | Power profiles | Centralized | Cross-Platform |
| 25 | (W. Zhang et al. 2014) | Performance, Deadline | Interference-aware | Centralized | Hadoop |
| 26 | (Hwang and Kim 2012) | Deadline, Cost | Pricing List, Bin Packing | Centralized | Hadoop |
| 27 | (Jyothi et al. 2016) | Performance, Deadline, Scalability | Job Profiles, Task Packing, Resource Reprovision | Centralized | Hadoop |
| 28 | (Guo et al. 2017) | Performance, Scalability | Slot Management, Speculative Execution | Centralized | Hadoop |
| 29 | (Orhean, Pop, and Raicu 2018) | Performance | Reinforcement Learning | Centralized | Cross-Platform |
| 30 | (Polo et al. 2011) | Performance, Scalability | Job Profiles, Slot reconfiguration | Centralized | Hadoop |
| 31 | (Kc and Anyanwu 2010) | Deadline | Greedy Heuristic | Centralized | Hadoop |

Now, each of the following subsections will provide a detailed study on the existing literature from the perspective of the scheduling objectives.

### 4.1.1  Performance-oriented Scheduling

Performance improvement in scheduling can be achieved from two levels. First one is from the application/job level; where the target is to minimize the execution time of a job. The second one is from the cluster level; where a cluster scheduler has a global goal to improve the performance of the whole cluster. The most optimized way of scheduling is doing both. First, the job performance can be modeled by building mathematical models, machine-learning models, using monitoring data, etc. to set the appropriate resource requirement and configuration parameters for a job which is enough to maintain its SLA. Then, while each job is submitted, the cluster level scheduler improves the performance of the job by various techniques such as task consolidations in the same node to reduce network transfers, placing tasks close to data, order jobs based on their priority or deadline, etc.

### 4.1.2  Cost-efficient Scheduling

The monetary cost of running a big data processing cluster in a cloud environment is crucial. Improper resource selection and resource scheduling might lead to resource wastage which intern increases the monetary cost of the cluster. If using VMs as the worker nodes of a big data cluster, it is often useful to turn-off unused or underutilized VMs to save cost if it does not affect performance/SLA of the jobs. Saving cost is mostly comes with a sacrifice of performance guarantee as cost can be saved by using a smaller number of resources in a cluster which might impact performance. Therefore, when both cost and performance is considered, resources are saved/consolidated only after ensuring a satisfying performance for all the jobs. In extremely scalable or fast scheduling systems, improving performance is the only goal and cost saving is mostly ignored.

### 4.1.3  Deadline-oriented Scheduling

Some jobs are associated with deadlines, and some job is time-critical or real-time and needs to be scheduled as soon as they arrive. Therefore, the deadline is an SLA parameter, and many schedulers try to minimize deadline violations. There are several techniques to achieve this — first, the pre-emption mechanism where non-priority jobs are killed when priority jobs need to be scheduled. Second, reserving some resources that can be dedicated to time-critical or deadline-constrained jobs only. Lastly, ordering the jobs beforehand based on their deadlines. However, maintain the job deadline while handling other SLA constraints for jobs is difficult due to the presence of stragglers (large periodic jobs that might hold a considerable chunk of resources), job inter-dependency (a deadline-constrained job might wait for other critical or non-priority jobs), etc. When multiple objects such as cost, deadline and performance are considered together, generally there are strict priorities between the objectives. For example, the first objective is always ensuring a satisfiable performance of a job so that it meets its given deadline. When these objectives are satisfied, only then cost-saving is considered.

### 4.1.4  Energy-efficient Scheduling

One of the significant challenges of running big data applications in cloud deployed cluster is minimizing their energy costs. Electricity used in the data centers in the USA accounted for about 2% of the total electricity usage of the whole country in 2010. Furthermore, each year, the energy consumption by data-centers is increasing at over 15%. Lastly, the energy costs can take up to 42% of a data-centers total operational cost. It is predicted by IDC (Internet Data Corporation) that by the year 2020, big data analytics market will surpass $200 billion. Therefore, more and more data-centers are made, and these data-centers will consume a tremendous amount of energy soon. Consequently, it is crucial to make the scheduling techniques energy-efficient from both the application and the cluster side. Furthermore, from both the cluster and application side, consolidating resources to save cost leads to energy saving as it helps to reduce the number of active physical machines from the infrastructure side.

### 4.1.5  Scalable Scheduling

Scalable scheduling means that the resource management or scheduling algorithms are scalable to large clusters and can perform in the presence of high number of scheduling requests in a heterogeneous environment. Although the centralized approach of scheduling is less complicated to handle the complex steps of scheduling at one place, it is not as scalable as a distributed/hybrid approach of scheduling. It can be observed that scheduler scalability is addressed in only a few works (2, 3, 4, 6, 11, 28, 30) which mostly have distributed/hybrid architecture. However, as the existing cluster systems are growing massively on size and scale to handle massive amounts of analytics demands, future research should focus on the distributed or hybrid deployment of schedulers to make them scalable.

## 4.2  Approach

The solution method towards the scheduling problem varies. Generally, a complete and sophisticated scheduler has separate performance prediction and resource assignment modules. The performance models are built from mathematical models to predict the runtime of a job, cost of running a job, deadline violation, etc. in advance which helps to make accurate scheduling decisions. Constraint programming-based approaches try to minimize or maximize an objective by satisfying the constraint parameters set by the job and the restrictions of resources on the cluster. However, for both resource assignment/allocation and scheduling, the optimization problem is always modeled as an NP-Hard problem. Therefore, even if exact algorithms or constraint solving approach can find optimal scheduling decisions, it is not feasible in most of the case and only applicable in small-scale clusters. In contrast, heuristics or meta-heuristics approaches are faster, less-complicated and provided acceptable near-optimal solutions and can be scalable to large clusters. Nowadays, machine learning approaches are also becoming popular to build sophisticated and intelligent schedulers.

## 4.3 Architecture

Some scheduling designs are centralized, some are distributed. Recently, some hybrid approaches have also been proposed which uses both a distributed or local scheduler and a global scheduler. Generally, there are two levels in scheduling. One is at the cluster manager level which manages and schedules all the jobs submitted from multiple users. Another one is on the application level that schedules the tasks of a job to the allocated resources by the cluster-level scheduler. A centralized cluster-level scheduler design is less complicated as it controls all the jobs. However, for a massive cluster, a centralized scheduler could be a single point of failure. This limitation is solved with either having backup master nodes with the cluster manager (using tools like ZooKeeper) or by designing a distributed scheduler where the worker nodes co-ordinate with each other to manage the tasks from different jobs.

## 4.4 Platform

Most of the researches have tried to design efficient scheduling algorithms for Hadoop MapReduce based clusters as it was the mostly used distributed data processing platform in the last decade. However, as Apache Spark, Apache Storm, etc. are becoming more popular and vastly replacing Hadoop these days, the researchers are focusing on these platforms now to devise scheduling algorithms. Lastly, due to the popularity of the cluster managers that support multiple different big data frameworks at the same time (Apache Mesos), or container-based platforms (Docker, Kubernetes); research has been going on building cross-platform cluster-level schedulers that can work with a cluster manager to effectively handle jobs from different platforms.

To summarize, it is always a hard challenge to provide a general scheduling strategy for all types of big data applications. To design a sophisticated scheduling algorithm, the type of big data application needs to be detected. Furthermore, depending on the user SLAs, the objectives should be chosen carefully. Then after setting the priority between different objectives, a suitable scheduling strategy can be devised.

# 5 Future Research Directions

### 1. Energy-efficient Fog/IoT deployment

Fog/IoT is going to become the most investigated area in the next decade because of the availability of a vast number of wearable devices, smartphones, smart sensors, etc. Therefore, the distributed deployment of data processing applications will be typical. However, it is not efficient to send all the data to process in the cloud data-centers as it might impose excessive network/transmission/bandwidth overhead in the whole system and increase the energy consumption of the data-centers. Therefore, energy-efficient software systems need to be developed that can process and analyze data on the edge/fog level to reduce energy consumption and boost the performance of time-critical applications. Also, it will help to meet the SLA requirement through multi-tiered resource management over the cloud data-center, fog nodes, and mobile devices.

### 2. Intelligent Resource Management

Machine learning algorithms are becoming more accurate and suitable for solving complex problems. Specifically, it is useful in resource management across all the different components. The resource usage statistics, system status, and the configuration parameters can be used to predict the system performance. Additionally, machine learning can be used for predicting anomaly, resource demand, peak usage period which will help to build sophisticated scheduling, resource scaling, and load-balancing algorithms. Lastly, small applications focusing on resource monitoring, performance analysis, local scheduling can be packaged as containers in a system that runs through a containerized management system to push small resource management components on the fog/edge level for achieving faster and flexible services.

### 3. Shared-sensing in IoT

As the number of IoT and mobile devices is increasing, a vast amount of resources from multiple users can be underutilized which neither energy-efficient nor cost-effective. Therefore, fog/IoT devices from various service providers and customers can be used collaboratively to provide efficient services. However, the software architecture should be made in such a way that it is both secure and beneficial for the collaborating partners. Besides, new protocols need to be designed on how to set the monetary cost and discount in a shared IoT infrastructure.

## 6  Summary

In this chapter, we have discussed the basics of cloud computing, the emergence of big data, processing platforms and tools used to handle big data applications and an overall view of resource management for big data applications in the cloud. We have specifically focused on the job scheduling aspect of resource management and provided a detailed taxonomy of job scheduling for big data applications. Furthermore, we have discussed the relevant research in scheduling and showed comparisons of various approaches regarding different aspects of scheduling. Lastly, we have highlighted some new research directions that need to be investigated to cope with the advanced resource management requirements in the modern era.